\documentstyle[epsf,rotate]{elsart} 
\begin{document}

\begin{frontmatter}

\title{On the definition of physical temperature and pressure for nonextensive thermostatistics}
\author{Ra\'ul Toral}  
\address{Instituto Mediterr\'aneo de Estudios Avanzados  
(IMEDEA), UIB--CSIC, \\Campus UIB,  
07071 Palma de Mallorca, Spain} 
\thanks[email]{e-mail: raul@imedea.uib.es} 
\thanks[web1]{URL: http://www.imedea.uib.es}

\begin{abstract} 
It is shown that a recent proposal to give physically meaningful definitions of temperature and pressure within Tsallis formalism for non-extensive thermostatistics leads to expressions which coincide with those obtained by using the standard Boltzmann formalism of Statistical Mechanics. 
\end{abstract}

\begin{keyword} 
Tsallis statistics. Microcanonical ensemble. \\
PACS: 05.20.-y, 0.5.20.Gd
\end{keyword}

\end{frontmatter}

In a recent paper\cite{ampp01}, Abe et al. have introduced so--called {\sl physical} definitions for the temperature, $T_{phys}$, and the pressure, $P_{phys}$ within the context of Tsallis generalized Statistical Mechanics. Their definitions are, respectively:
\begin{equation}\label{tphys}
T_{phys}=\left(1+\frac{1-q}{k}S_q\right)\left(\frac{\partial S_q}{\partial U_q}\right)^{-1}
\end{equation}
and
\begin{equation}\label{pphys}
P_{phys}=\frac{T_{phys}}{1+\frac{1-q}{k}S_q}\left(\frac{\partial S_q}{\partial V}\right).
\end{equation}
Here, $S_q$ is the Tsallis entropy and $U_q$ the internal energy. $V$ is the system volume and $k$ is a constant. The index $q$ measures the degree of non-extensivity of the entropy in the sense that, for two independent subsystems $A$ and $B$, the entropy satisfies:
\begin{equation}\label{sqab}
S_q(A+B)=S_q(A)+S_q(B)+\frac{1-q}{k}S_q(A)S_q(B).
\end{equation}

I will show that the above expressions for the physical temperature and pressure coincide with the ones given by the standard application of the microcanonical ensemble. Therefore, it turns out that, when rewritten in term of these physical parameters, the results obtained by application of Tsallis statistics coincide with the ones derived by application of the usual (Boltzmann's) formalism of Statistical Mechanics.

Let me begin with the definition of entropy given in Tsallis original paper\cite{t88}:

\begin{equation}
S_q= k \frac {1- \sum_{i=1}^W p_i^q}{q-1},
\end{equation}

where the $p_i$'s are a set of microscopic probabilities given to each of the $W$ configurations defining a statistical ensemble. These probabilities are obtained by maximizing the entropy $S_q$ subjected to the appropriate constraints. In the microcanonical ensemble of constant energy, $U$, the maximization procedure leads to equiprobability:

\begin{equation}
p_i=\left\{\begin{array}{ll}\Omega(U)^{-1}, &  \varepsilon_i = U \\
0, & {\rm otherwise.} \end{array}\right. 
\end{equation}

Where $\varepsilon_i$ is the energy of the configuration whose probability is $p_i$, and $\Omega(U)$ is the number of configurations having a given energy $U$. It follows readily that the entropy can be expressed in terms of $\Omega(U)$ as:

\begin{equation}\label{squ}
S_q(U)= k\frac {1-\Omega(U)^{1-q}}{q-1}.
\end{equation}

In the case $q=1$, this expression tends to the result $S(U)=k\ln \Omega(U)$. Rewriting Eq. (\ref{tphys}) as

\begin{equation}
\frac{1}{k T_{phys}}= \frac{1}{1-q}\frac{\partial  \ln[1+\frac{1-q}{k}S_q]}{\partial U_q}
\end{equation}

and replacing $S_q$ as given by Eq. (\ref{squ}), identifying the internal energy $U_q$ with the constant system energy $U$, one gets 
\begin{equation}
\frac{1}{k T_{phys}}=\frac{1}{1-q}\frac{\partial \ln \Omega(U)^{1-q}}{\partial U},
\end{equation}
which, after simplification, becomes Boltzmann's relation:
\begin{equation}\label{physt}
\frac{1}{k T_{phys}}=\frac{\partial \ln \Omega(U)}{\partial U}.
\end{equation}
It is not surprising that one obtains precisely this relation between the physical temperature $T_{phys}$ and the number of microscopic configurations $\Omega(U)$. On one hand, Abe et al. derive their result by considering two independent subsystems, $A$ and $B$ in contact with each other, such that the total internal energy is fixed: $U(A+B)=U(A)+U(B)$, and thermal equilibrium is characterized by the maximum total entropy state. On the other hand, the standard derivation of (\ref{physt}), maximizes the  number of configurations of the composite system $\Omega(A+B)$ assuming\cite{pathria} that it satisfies $\Omega(A+B)=\Omega(A)\Omega(B)$. Since, within the Tsallis formalism, this equality follows from (\ref{sqab}) and (\ref{squ}), it is clear that both approaches should give the same answer.

Similarly, for the physical pressure, replacing (\ref{squ}) in (\ref{pphys}) it is straightforward to obtain the relation:
\begin{equation}
P_{phys}=kT_{phys}\frac{\partial \ln \Omega(U)}{\partial V},
\end{equation}
which coincides with the definition of the pressure in the microcanonical ensemble of Boltzmann\cite{pathria}.

Finally, the defition of free energy $F_q$ given in \cite{ampp01}:
\begin{equation}
F_q=U_q-T_{phys}\frac{k}{1-q}\ln\left(1+\frac{1-q}{k}S_q\right),
\end{equation}
becomes, after replacing Eq.(\ref{squ}), and identifying the internal energy $U_q$ with the system energy $U$:
\begin{equation}
F_q=U-k T_{phys}\ln \Omega(U),
\end{equation}
which again coincides with the standard definition of free energy in the microcanonical ensemble\cite{pathria}. 

Notice that the number of configurations with a given energy $\Omega(U)$ depends only on the system Hamiltonian and does not rely on any {\em a priori} assumption on the actual relation of the entropy to the configuration probabilities $p_i$. Therefore, all the dependence on the $q$ parameter disappears when computing thermodynamical properties (such as the free energy) in terms of the {\em physical} temperature and pressure using the microcanonical ensemble. Assuming equivalence between statistical ensembles\footnote{This equivalence is ultimately needed to justify the use of ensembles\cite{pathria} and has been shown explicitely by a direct numerical calculation of the number of states $\Omega(U)$ in short and long-range Ising models\cite{st01}, for $q \le 1$ (the physically relevant case, see \cite{ampp01})}, the results shown in this paper explain in a simple way why some well established results of Statistical Mechanics, such as the equation of state and specific heat for ideal gases, equipartition and virial theorems, etc. are being reproduced by some recent calculations\cite{ampp01,mpp00} within the framework of Tsallis statistics independently of the value of $q$.

\noindent Acknowledgments

I thank C. Tsallis, S. Abe and A.R. Plastino for several discussions about
the Tsallis statistics. Financial support from DGESIC (Spain), projects
PB97-0141-C02-01 and BFM2000-1108, is acknowledged.

\end{document}